\documentclass[aps,prl,twocolumn,superscriptaddress,showpacs]{revtex4}
\usepackage{graphicx}

\newcommand{\ba}{\begin{eqnarray}}
\newcommand{\ea}{\end{eqnarray}}
\newcommand{\beq}{\begin{equation}}
\newcommand{\eeq}{\end{equation}}
\newcommand{\be}{\begin{equation}}
\newcommand{\ee}{\end{equation}}

\begin{document}
\title{Nuclear binding, correlations and the origin of EMC effect}

\author{Omar Benhar}
\email{benhar@roma1.infn.it}
\affiliation{INFN, Sezione di Roma, I-00185 Roma, Italy}
\affiliation{Dipartimento di Fisica,  ``Sapienza'' Universit\`a di Roma, 
I-00185 Roma, Italy}
\author{Ingo Sick}
\email{ingo.sick@unibas.ch}
\affiliation{Dept. f{\"u}r Physik, Universit{\"a}t Basel, 
 CH-4056 Basel, Switzerland} 
\vspace*{5mm}
\begin{abstract}
\vspace{3mm}
%%% NEW replace the following abstract by the new abstact
%Recent data  for {\em light} nuclei, with   $3 \leq A \leq 12$, have the potential 
% to shed new light on the origin of the EMC effect in the  
%intermediate-$x$ region. Here we study the role of nuclear binding using the understanding of
%the average nucleon removal energies, $\bar{E}$, provided by state-of-the-art calculations 
%based on nuclear many body theory. We find an excellent correlation between
%the EMC data at intermediate $x$ and $\bar{E}$ for nuclei with $A$ from
%$3$ to $\infty$, indicating  that in this $x$ region binding is the 
%dominating variable driving the EMC effect. The role played by nucleon-nucleon 
%correlations in this context is also discussed.
%%% NEW: new abstract
Recent data  for the slope of the EMC-ratio in the intermediate $x$-region for {\em light} 
nuclei, with   $3 \leq A \leq 12$, have the potential  to shed new light on the origin of 
the EMC effect. Here we study the role of nuclear binding using the scaling variable ${\tilde y}$, best suited to take into account this effect, and the understanding of
the average nucleon removal energies, $\bar{E}$, provided by state-of-the-art calculations 
based on nuclear many body theory. We find an excellent correlation between
the new EMC data at $x \sim 0.5$ and $\bar{E}$ for nuclei with $A$ from
$3$ to $\infty$, indicating  that in this $x$ region binding is an important ingredient to
explain the EMC effect. The role played by nucleon-nucleon 
correlations in this context is also discussed.
%%% CHANGE: more explicit on the observable (slope), more precise on x-region, less 
%%% strong claim ('important ingredient' instead of 'driving'
%%% END new abstract
\end{abstract}

\pacs{25.30.Fj, 24.85+p, 13.60Hb, 21.60De}

\date{\today}
\maketitle

The so called {\em EMC effect} was, in 1983, a surprising discovery 
\cite{Aubert83a}.
The ratio $R(x)$ between the Deep Inelastic Scattering (DIS) cross sections of
leptons on iron  and the deuteron 
was found to sizably differ from unity. At 
low values of the Bjorken scaling variable $x$ the ratio is $<1$, at medium values 
of $x$ $R$ drops from $\sim 1$ to values as low as 0.8 and at large $x$ it reaches 
values larger than 1. While the latter feature can be quantitatively explained by 
the smearing of the parton distribution functions arising from the momentum 
distribution of nucleons in nuclei, the former one is accounted for by including 
the effect of nuclear shadowing.  The reduction of $R(x)$ at intermediate $x$ 
was not easily explained, although many different models have been proposed 
(for reviews of the extensive literature see, e.g., Refs.~\cite{Norton03,Arneodo94,
Geesaman95}). Often, but not always, these models do involve the role of the 
binding of nucleons in the nucleus. In general, however, the effect of binding 
alone is not large enough to reproduce the data \cite{Ciofi91}. 
%***
% since we want to protect ourselves from the request to compute the slope
%using the $\tilde{y}$-approach, it might be useful to add here a sentence
%concerning pions and FSI. I added one.
%***
In particular, the rise towards small $x$ seems to
require the contribution  of  pions --- enhanced due to nucleon binding --- \cite{Kulagin06}
and consideration of the role of the recoil final state interaction. 
No ``definitive'' explanation has surfaced, in part due to the fact that no model
provides a quantitative description of the whole range of $x$, unless very specific assumptions
are made.

Recently, new measurements have been performed by Seely {\em et al.} \cite{Seely09} 
with the hope to elucidate things better by adding precise data on {\em light} nuclei.
For these nuclei, the nuclear structure is extremely well understood, 
mainly thanks to Green's Function Monte Carlo (GFMC) calculations based on 
state-of-the-art models of the nuclear hamiltonian, strongly constrained by the 
observed properties of the two- and three-nucleon systems. For light nuclei, 
nuclear properties change rapidly as a function of the mass number
$A$, so that different models for the EMC effect could yield rather different 
predictions. The data of Ref.~\cite{Seely09} mainly cover the intermediate region of 
$0.35 \leq x \leq 0.7$, 
which is little affected by smearing and away from the low-$x$ region, where the
potential contributions of pions, coherence effects  and final state 
interactions complicate matters. 

In order to provide data least affected by experimental errors, the authors of
Ref.~\cite{Seely09} have studied in particular the slope $dR/dx$, which is
insensitive to normalization errors. The correlation between
$dR/dx$ for A=3,4,9,12 and   the average density \cite{Seely09} has been 
studied in order to find a phenomenological relation that could further elucidate the
EMC-effect at mid-$x$; no unambiguous ($\pm$linear) correlation has been found. In 
particular, the $dR/dx$ value for $^9$Be does not follow the tendency observed 
for the other light nuclei. The relation to quantities such as nuclear binding
energy or nucleon separation energy \cite{Fomin11} also has produced no further
insight. A correlation with the  inclusive quasi-elastic cross section for 
electron scattering at  $x > 1$ has been found \cite{Weinstein11,Hen12}, and 
attributed to the role of nuclear high-momentum 
components that increase with increasing $A$. 
However, in the standard treatment of Fermi motion, folding the nucleon parton 
distributions with nuclear momentum distributions alone would lead to values 
of the EMC ratios $>1$ and a mid-$x$ slope $dR/dx$ of the wrong sign. 

In this paper, we argue that DIS data are best analyzed in terms of the scaling variable 
${\tilde y}$, widely employed in studies of a variety of scattering processes involving 
composite targets. We explore the correlation of $dR/d{\tilde y}$ with nuclear binding, which 
in many approaches is advocated as an important element for the 
explanation of the EMC effect, and is responsible in particular for the decrease of the
EMC-ratios to values $<1$ at intermediate values of the scaling variable. 

%%% NEW. Replace the following paragraph
%In general,  DIS cross sections are studied in the framework of 
%the parton model, which explains the occurrence of scaling and provides the 
%basis for the determination of the parton distribution functions \cite{Bjorken69}. 
%The constituents the lepton scatters from are assumed to be point-like
%spin one half particles carrying, in the Infinite Momentum Frame (IMF), 
%a fraction $x$ of the nucleon four-momentum. 
%This quantity  is identified with the Bjorken scaling variable $x=Q^2/(2M\nu)$, where $Q^2$,
% M and $\nu$ are the squared four-momentum transfer, the nucleon mass and the lepton
% energy loss, respectively.  Within this approach  the introduction of binding 
%(off-shellness) of nucleons in nuclei is conceptually difficult, as partons in the
%nucleon are intrinsically on-shell. 
%%% NEW paragraph begins. Omar note that I have placed the "conceptually.." at the end.
In general,  DIS cross sections are studied in the framework of 
the parton model, which explains the occurrence of scaling and provides the 
basis for the determination of the parton distribution functions \cite{Bjorken69}. 
The constituents the lepton scatters from are assumed to be point-like
spin one half particles carrying, in the Infinite Momentum Frame (IMF), 
a fraction $x$ of the nucleon four-momentum. 
This quantity is identified with the Bjorken scaling variable $x=Q^2/(2M\nu)$, 
where $Q^2$, M and $\nu$ are the squared four-momentum transfer, the nucleon mass and 
the lepton energy loss, respectively.   In the IMF
%*** 
%The statement on the vanishing FSI effects in the IMF is wrong, as we know from the
%work of Brodsky, Hoyer.... . (Do you think we need a reference?)
% We therefore should be more careful and  say "thought to vanish"
%***
interaction effects are expected to vanish 
%due to relativistic time dilation 
\cite{Feynman72}, and the partons in the nucleon are on-shell. Within this approach  the 
 binding  (off-shellness) of nucleons in nuclei is conceptually difficult to introduce.
%%% NEW. paragraph ends

%This difficulty is reflected in the diversity of 
%approaches different authors have followed  to include nucleon binding. 

In order to avoid this  problem, we here employ a somewhat 
different approach, first developed to analyze quasi-elastic electron-nucleus 
scattering, a process perfectly analogous to DIS. Scaling of the quasi-elastic 
scattering  cross sections has been studied in terms of 
the scaling variable $y$ \cite{Sick80} which is derived from the kinematics of
the underlying process of elastic scattering from nucleons bound in nuclei. 

%*** 
%Below, we should avoid the confusion that, in the case of $\tilde{y}$, we
%neglect the nucleon mass M. I therefore would prefer to write eqs.(1,2) using a
%generic mass (little) m, explaining that for the case of quasi-elastic
%scattering m=M. I have made the corresponding change in eq.(1) and below. Also,
%I think it is better to mention immediately after the first reference to
%"off-shell" what off-shell means. In order not to break to flux of ideas, I 
%therefore would propose that the text on the "interaction energy ...." be 
%moved to after the explanation of "off-shell". I made the corresponding change. 
%I have also, after explaining that there is no $\sqrt{|{\bf k}|^2+m^2}-m$  term,
%added that, instead there is a term $m$. For consistency of notation, I
%capitalized $M_u, M_d$
%***

Energy-momentum 
conservation of elastic scattering from a constituent of mass $m$ 
$\bf{k}$ yields
\ba 
\label{e:conservation}
\nu = \sqrt{|{\bf k}+{\bf q}|^2+m^2}-m
\ea
with $\bf{q}$ being the 3-momentum transfer.  
and $m$ being, for quasi-elastic scattering,  the mass of the nucleon, {\em
i.e.} $m=M$.
Note that in this approach the initially bound nucleon is assumed to be
{\em off-shell}.
%; there is no $\sqrt{|{\bf k}|^2+m^2}-m$ term on the right-hand 
%side.  
%%% NEW. Add the following 3 lines
Off-shell means that there is no $-\sqrt{|{\bf k}|^2 + m^2}$ term on the right-hand 
side, only a term $m$; while the nucleon has momentum $\bf{k}$ it is assumed 
to be {\em bound} with total
energy =0. As usual, the interaction energy of the knocked-out nucleon with
 momentum  $\bf{k}+\bf{q}$ is neglected, which at large  $|\bf{k}+\bf{q}|$ 
in general is a good approximation.
%%% END NEW
 An additional term 
%%% NEW: in the following line replace "often" by "normally"
$\bar{E}>0$, accounting for the average nucleon removal energy, is  normally
added on the right-hand side of Eq.(\ref{e:conservation}). 

Neglecting the component of ${\bf k}$ perpendicular to $\bf{q}$, 
which is justified in the
limit of large $|{\bf q}|=q$, yields the standard scaling variable introduced in Ref.~\cite{Sick80}
\ba
\label{def:y}
y =  \sqrt{\nu^2+2 m \nu} - q \ .
\ea  
The physical meaning of $y$ is straightforward: it is the component of the 
momentum of the initially bound nucleon parallel to $\bf{q}$ in the rest frame
of the nucleus. 

Here, we specialize $y$ to the conditions appropriate for DIS on the nucleon, i.e. 
scattering from (basically) up- and down-quarks, the  
rest masses  of which, $m = M_u, M_d <$10~MeV,  can be safely neglected at the 
energies 
relevant to DIS. In this case, the expression for $y$ of Eq.(\ref{def:y}) simplifies to  
\ba
\tilde{y} = \nu - q \ ,
\ea
where a 'tilde' has been added to remind ourselves that $\tilde{y}$ corresponds
to the $m \to 0$ limit \cite{Benhar00}. The physical meaning of $\tilde{y}$ is
 analogous to the one of $y$ in quasi-elastic electron-nucleus scattering: $\tilde{y}$ 
is the component of the $u$/$d$-quark momentum  parallel to  $\bf{q}$ in the rest
frame of the nucleon.

%%% NEW: replace the following paragraph
%Note that, as the $u$/$d$ constituents are assumed to be off-shell in the nucleon, the 
%additional off-shellness due to the binding of the nucleon in the nucleus can 
%be introduced without conceptual difficulties. Since in the nuclear DIS process the hit 
%nucleon  --- or, more precisely, the  recoiling quark and the remaining 
%debris of the  nucleon --- are removed from the 
%nucleus, the process requires an extra energy equal to the mean removal energy 
%$\bar{E}$  
%of a nucleon, which the scattered lepton's energy loss has to provide. Hence,  
%for DIS from nuclear targets 
%\ba
%\tilde{y} = \nu  - \bar{E} -q \ .
%\ea
%%% NEW: new paragraph
%*** 
%I added below "on nuclei"
%***
Due to the large energy and momentum transfer in DIS on nuclei, the system of hit quark plus
remaining debris of an initially bound nucleon will leave the nucleus with 
high energy and momentum, and
the interaction with the (A--1)-nucleus is weak.  The removal of this system costs an
energy equal to the nucleon mean removal energy $\bar{E}$ which the scattered lepton 
 has to provide. For the DIS process on the nucleon proper only the 
 energy $\nu^\prime = \nu - \bar{E}$ is available. This leads to  
\ba
\tilde{y} = \nu^\prime - q = \nu  - \bar{E} -q \ .
\ea
%*** 
%To make above eq. more readable, if have switched around $\bar{E}$ and q
%*** 
As both the energy of the quark in the nucleon and $\bar{E}$ are well defined quantities in the rest 
frame of the nucleon (nucleus) they can be added without conceptual difficulties. 
%%% NEW: end of the new paragraph. 
%%% I am not sure whether the last sentence is better than 
%%% the first sentence of the old paragraph:"Note that..."
 
For the isolated nucleon, {\em i.e.} for $\bar{E}=0$, using the variable $\tilde{y}$ is known to 
leads to a  scaling of the DIS cross section which is even better in quality than the 
scaling observed in terms of Bjorken $x$. 
%***
%It would perhaps be better to say "can be shown easily" instead of "should not
%be surprising". I changed it.
%*** 
This can be shown easily, as  
$\tilde{y}$ is trivially related to the Nachtmann variable $\xi$ 
\cite{Nachtmann73}. While being usually defined using the much less transparent equation
\ba
\label{def:csi}
\xi = 2x/(1+\sqrt{1+4M^2 x^2/Q^2}) \ ,
\ea
Nachtmann's $\xi$ can easily be shown to reduce to $(q-\nu)/M$, implying 
$\tilde{y} = -\xi M $. 
Using Nachtmann's $\xi$, which becomes identical to $x$ in the $Q^2/\nu^2 \to 0$ 
limit, is known to extend the scaling property of DIS to lower $\nu$. 

As compared to $x$ or $\xi$,  $\tilde{y}$ has a well defined physical 
meaning \footnote{Although $\xi$ is often attributed a physics meaning by 
claiming that, as compared to $x$, it includes ''target mass corrections'', 
the  equation $\xi  = -(\nu -q)/M$
shows that, as for $x$, the target mass $M$ only serves to make the
variable dimensionless.} in the nucleon rest frame, the coordinate system where 
the DIS experiments are usually done and where theoretical studies, such as 
lattice QCD calculations, hope to provide a quantitative understanding of the 
nucleon structure functions 
%{\bf we may provide a reference here. What do you think?}.

In order to calculate $\bar{E}$  directly 
one would need to know the spectral function $S(k,E)$, giving the probability 
to find in the nucleus a nucleon with momentum $k$ and removal energy $E$.  

Theoretical calculations of the spectral function require the knowledge of both 
the $A$-nucleon ground state and the full spectrum of eigenstates of the $(A-1)$-nucleon 
system. In addition, they involve an intrinsic degree of complexity, associated with the evaluation 
of non diagonal nuclear matrix elements, that rapidly increases with $A$. 

Studies based on nuclear many body theory and 
realistic hamiltonians have been carried out for A=3 
\cite{Dieperink76,Ciofi80,Meier83} and for isospin symmetric nuclear matter
\cite{Benhar89,Ramos89}, while in the case of nuclei with $A>12$ approximate 
spectral functions obtained using the local density approximation are available 
\cite{Sick94}.  In the present work, aimed at studying the correlation between 
binding effects and the EMC effect for $3\leq A \leq 12$, we exploit 
Koltun's sum rule \cite{Koltun72,Bernheim72} to obtain the average removal energies
from the results of GFMC calculations. 

In the absence of three-nucleon interactions, Koltun's sum rule states that
\ba
\label{def:koltun}
%%% NEW: in the line below the last "+" should be replaced by a "-"
\frac{E_0}{A} = \frac{1}{2} \, [ \,{\bar T} \frac{A-2}{A-1} - {\bar E} \, ] \ ,
\ea
where $E_0/A$ is the nuclear binding energy per particle obtained from nuclear
masses and includes a (small) correction for the Coulomb energy,
\ba
\label{def:average}
{\bar T} = \int d^3k dE \ \frac{k^2}{2M} S(k,E) \  , 
\ea
and
\ba
{\bar E} = \int d^3k dE \  E \  S(k,E) \ .
\ea
The small contribution of the three-nucleon potential $V_3$, which is known to 
be needed to achieve a precise determination of $E_0$ for $A \geq 3$, can be 
taken into account \cite{Harada76} by adding a term  $\langle V_3 \rangle$/A,  where 
$\langle \ldots \rangle$ denotes the ground state expectation value, to the 
right hand side of Eq.(\ref{def:koltun}).

It has to be pointed out that the Koltun sum rule is an {\em exact} result, although 
its experimental verification involves severe difficulties. The results of the 
analysis of the $(e,e^\prime p)$ reaction carried out by Bernheim {\em et al.}
\cite{Bernheim72}, 
suggesting that the sum rule is badly violated, are due to  the 
limited kinematical range covered by the experiment, which does not include 
the contribution of correlation effects, leading to the appearance of tails of 
$S(k,E)$ extending to large energy and momentum \cite{Rohe04c}.

The GFMC approach allows one to obtain essentially exact binding energies for the 
ground states and very good estimates of the energies of low-lying excited states. 
The wave functions resulting from GFMC calculations have also been employed to 
obtain density and momentum distributions, electromagnetic form factors and 
spectroscopic factors, as well as to compute many electroweak processes of 
astrophysical interest. These studies have clearly shown that the {\em ab initio} 
approach based on the numerical solution of the many body Scr\"odinger equation 
and modern nucleon-nucleon interactions, fitted to NN scattering data, is capable 
to 
describe the full complexity of nuclear structure, including single particle 
properties, correlations and clustering. 

For all the light nuclei $A \leq 12$ of interest here, GFMC calculations have 
been carried 
out [for the lighter nuclei, exact solutions of the Schr\"odinger equation 
are also available from simpler approaches, such as Variational Monte Carlo (VMC)]. 
From the resulting binding energies, 
momentum distributions  and (small) 3-body contributions we have calculated  the
average removal energies $\bar{E}$ which are listed in the following table.    
\begin{center}
\begin{table}[htb]
\begin{tabular}{l|r|c|c}
nucleus & removal~ & method & reference  \\
        & energy~~ &        &            \\[1mm]
\hline
\rule{0mm}{5mm}
~$^2$H~  & 2.2 MeV  & & \cite{Schiavilla86} \\[1mm]
~$^3$He & 14.6 MeV & GFMC & \cite{Pieper01} \\[1mm]
~$^4$He & 35.8 MeV & VMC & \cite{Pieper01} \\[1mm]
~$^9$Be & 43.8 MeV & GFMC & \cite{Pieper07} \\[1mm]
$^{12}$C & 52.2 MeV & GFMC & \cite{Pieper07} \\[1mm]
~NM     & 70.5 MeV & FHNC  & \cite{Akmal98} \\[1mm]
\hline
\end{tabular}
\caption{ Average  removal energies used in this work.} 
\end{table}
\end{center}
The values of $\bar{E}$ are significantly larger than the typical values that 
have been used in previous studies of the EMC effect \footnote{Early
exceptions are Refs.~\cite{Dieperink91,Ciofi91}.}, 
studies  which often 
%%% NEW replace in the line below "effects" by "effects alone"
found that binding effects alone were not large enough to explain the EMC data. 
For $^{12}C$,  for example, the average removal energy one would derive from 
the centroids of the s- and p-shell
removal energies as measured by $(e,e'p)$ experiments \cite{Mougey76} is of the 
order 25 MeV. A similar value is obtained from mean-field calculations, such 
as Hartree-Fock. The
difference to the value listed in Table 1, 52~MeV,  is due to the fact that the 
s- and p-shell peaks \cite{Mougey76} account only for (the $\sim$75\% of) 
%%% NEW: add below, after "orbitals" "\cite{Pandharipande97}" (the RMP Colloquium paper)
nucleons  occupying mean-field  orbitals \cite{Pandharipande97}. The short-range
correlations between nucleons, induced by the strong short-range components
of the NN interaction (both central and tensor) lead to the appearance of 
nucleons in states of high momentum and high removal energy; the corresponding 
strength, which is thinly spread over a
large range of initial momenta $k$ and removal energies $E$, gives an 
important contribution to $\bar{E}$. This picture has been confirmed by
$(e,e'p)$ experiments designed to provide a measurement of the 
spectral function  $S(k,E)$ at large values of $k$ and $E$ \cite{Rohe04c}.

Coming back to the EMC-effect, we show in Fig. 1 the correlation between the
%%% NEW: in the line below $dR/dy$ should be  $dR/d\tilde{y}$ 
average removal energy $\bar{E}$ and the derivatives $dR/d\tilde{y}$ determined from the
$dR/dx$ values of  Ref.~\cite{Seely09} 
%%% NEW: the following footnote can be put in if we have space
 \footnote{$\xi = -\tilde{y}/M$ differs by less than
5\% from $x$ and is calculated using {\em e.g.} eq.(\ref{def:csi}) and $\bar{Q}^2$ from 
\cite{Seely09}}.
%%% NEW: end footnote 
%  
\begin{figure}[hbt]                                                                
\begin{center}                                                                     
\includegraphics[scale=0.45,clip]{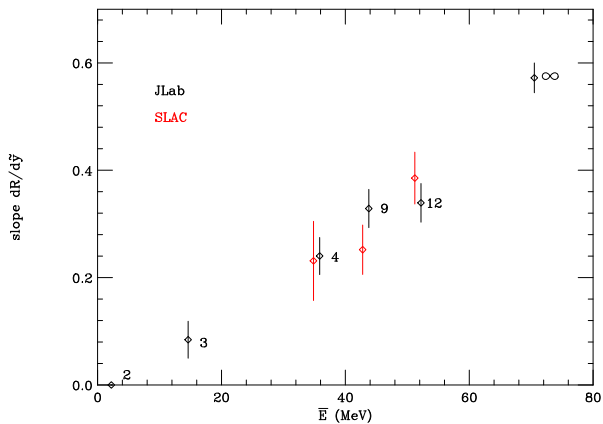}                                       
{\caption[]{Correlation between the average nucleon removal energy $\bar{E}$ 
%%% NEW in the line below $dR/dy$ should be  $dR/d\tilde{y}$ 
and the slope of the EMC-ratio $dR/d\tilde{y}$ obtained from the $dR/dx$ values of  
\cite{Seely09,Day89} at $x \sim 0.5$. 
Data points are labeled by the nuclear mass number A.}}                                                                   
\end{center}                                                                       
\end{figure}                                                                       
As Fig. 1 shows, there is an excellent linear correlation between $\bar{E}$ and
%%% NEW: in the line below $dR/dy$ should be  $dR/d\tilde{y}$  *TWICE*
$dR/d\tilde{y}$. This correlation is much better than the one of $dR/d\tilde{y}$ ($dR/dx$)
 with other
%*** One argument of the referees was that our conclusion is too strong. I
%therefore would prefer below to replace "tells us" by "indicates". I changed it.
%***
quantities (referred to above). This indicates that the driving quantity for
the EMC-effect at mid-$x$ values is indeed the binding of nucleons, as described
by the mean removal energy.

%%% NEW paragraph, to be added
To the extent that quasi-elastic electron-nucleus scattering at large $x$ could be 
identified with high-momentum components, the correlation with $a_2$ 
found in \cite{Weinstein11,Hen12} actually
would be  based on the same physics as discussed here. There is a strong correlation 
between $\bar{E}$ and $\bar{T}$ (for $^{12}C$, for instance, $\bar{E}$=25.2MeV, 
$\bar{T}$=31.2MeV, the difference leading to a comparatively small  $E_0/A$=6.1MeV). 
But while high-$k$ nucleons alone lead to EMC-ratios $R>1$ and a
positive slope near $x \sim 0.5$, the binding leads to a much larger effect in the opposite
direction, producing overall the $R<1$ and $dR/dx < 1$ as found by experiment.   
%%% END new paragraph

In Figure 1 we have also included the point corresponding to uniform, isospin symmetric, 
nuclear matter (NM). The average removal energy has been determined as discussed above from 
the variational results of Ref.~\cite{Akmal98}, obtained using the Fermi Hyper-Netted Chain (FHNC) summation 
%%% NEW in the line below $dR/dy$ should be  $dR/d\tilde{y}$ 
scheme. The slope $dR/d\tilde{y}$ has been extracted from 
Ref.~\cite{Day89} \footnote{We have added  Coulomb corrections
 \cite{Aste04b} neglected in \cite{Day89}; they change the slope by 
$\sim$ 2\%.}. 

As the determination of $R(x)$ for nuclear matter involved a
fit to the {\em world} EMC-data for all nuclei with mass number $A \geq 12$, 
the NM point 
is indicative of the fact that the excellent correlation between $dR/d\tilde{y}$ and
$\bar{E}$ is valid for {\em all} nuclei.  

We note that the correlation between $dR/dx$ and $\bar{E}$ (not shown) is very
similar to the one observed in Fig. 1. Only  the NM data point, which
on average corresponds to  larger $Q^2$ than the data of \cite{Seely09}, would
be slightly shifted due to a different conversion factor between $x$ and
$\tilde{y}$.
%***
%Here we could add a explanatory sentence, which I have added below.
%***
While numerically the difference between $x$ and $\tilde{y}$ is small, these
quantities differ radically in their physical interpretation.

%%% NEW: replace the paragraph below
%The approach based on many body theory employed in our work, while not 
%including some of the mechanisms which are believed to determine the 
%low-$x$ behavior of EMC ratio,  may be used to obtain theoretical estimates of 
%its slope at mid $x$. However, this would require the spectral 
%functions of nuclei with $4 \leq A \leq 12$, which are not available.
%%% NEW: start new paragraph
In principle, the approach based on many body theory employed in our work, while not 
including some of the mechanisms which are believed to determine the 
low-$x$ behavior of EMC ratio,  may be used to obtain theoretical estimates of 
its slope at mid $x$. However, to achieve the level of accuracy required for 
a meaningful comparison with the data, one would need spectral functions 
computed using the Green's function Monte Carlo technique and including 
the full set of eigenstates of the recoiling nucleus, which are 
not yet available.
%%% NEW: end new paragraph

In the case of infinite nuclear matter, $S(k,E)$ has been computed within the 
FHNC/SOC summation scheme, including the contributions of one hole and 
two hole-one particle states \cite{Benhar89}. However, compared to the 
calculation of the ground state expectation value of the hamiltonian 
discussed in Ref.~\cite{Akmal98}, the work of Ref.~\cite{Benhar89} 
was based on a oversimplified treatment of the
three-nucleon interactions and involved a number of additional technical difficulties 
(e.g. the orthogonalization of correlated states), leading to 
a larger theoretical uncertainty. 
As a result, the value of the average removal energy obtained
from the spectral function of Ref.~\cite{Benhar89}, ${\bar E} \approx 61 \ {\rm MeV}$, 
appreciably differs from the one reported in Table 1. 

In conclusion, we have shown that there is a strong correlation between the
EMC-effect at mid-$x$ and the average nucleon removal energy. This
correlation covers all nuclei, from $^3He$ all to way to infinite nuclear
%%% NEW: softer formulation: replace below "mainly drives" by "is important for"
matter. This confirms that the  binding of the nucleons in the 
nucleus is very important for the EMC effect at $x \sim 0.5$.

As binding plays an important role, our study of the data was done 
in terms of the scaling variable $\tilde{y}$. 
Being derived as  a property of initially off-shell quarks in the Lab frame, 
$\tilde{y}$ can be generalized to take into account the additional off-shellness due 
to nuclear binding without conceptual difficulties.
%***
%the repeated reference to NMBT breaks somewhat the flow of ideas concerning
%$\tilde{y}$. I therefore would propose to omit the ", by using
%the results of nuclear many body theory.". I have removed it.
%***
 It should also be emphasized that, 
while $\tilde{y}$ is
particularly suited to discuss binding effects,  it may also be  preferable to
Bjorken-$x$ in general as a scaling variable. 
Not only does $\tilde{y}$ yield better scaling,  it     
also  has a more intuitive physics interpretation as the
momentum component of the $u$,$d$-quarks parallel to $\bf{q}$ in the rest frame of
the nucleon, and allows for a unified description of inclusive scattering. 
%%% NEW: start new text
The analogous occurrence of $y$-scaling, and its straightforward interpretation, 
have  been exploited to extract valuable information from the analysis of a variety of 
scattering processes, ranging from photon scattering from electrons bound in atoms
 to neutron scattering from  quantum liquids  and quasi-elastic electron-nucleus
scattering \cite{Benhar08a}.
%%% NEW: end new text

We thank Robert B.  Wiringa for providing the VMC/GFMC momentum 
distributions and three-body force contributions employed to obtain the results
of Table 1.

%\bibliographystyle{unsrt}
%\bibliography{../sum2}

%\bibliography{/usr/users/sick/sum3}

\end{document}